\documentclass[showpacs,amsmath,amstex,amssymb,mathfonts,prb,twocolumn]{revtex4-1}
\usepackage{graphicx,bm,units}
\usepackage{amsmath,amssymb}
\usepackage{color, palatino,  wrapfig}

\newcommand{\ra}{{\rightarrow}}
\newcommand{\be}{\begin{equation}}
\newcommand{\ee}{\end{equation}}

\newcommand{\ba}{\begin{eqnarray}}
\newcommand{\ea}{\end{eqnarray}}

\newcommand{\ignore}[1]{}
\newcommand{\pb}{{\textbf{p}}}
\newcommand{\pdag}{\phantom{\dagger}}
\newcommand{\rb}{{\textbf{r}}}

\begin{document}

\title{Vortex Lines in Topological Insulator-Superconductor Heterostructures}

\author{Ching-Kai Chiu$^{1}$}\author{Matthew J.  Gilbert$^{2,3}$}\author{Taylor L. Hughes$^{1}$}
\affiliation{$^1$Department of Physics, University of Illinois, 1110 West Green St, Urbana IL 61801}
\affiliation{$^2$Department of Electrical and Computer Engineering, University of Illinois, 1406 West Green St, Urbana IL 61801}
\affiliation{$^3$Micro and Nanotechnology Laboratory, University of Illinois, 208 N. Wright St, Urbana IL 61801}
\date{\today}
\begin{abstract}
3D topological insulator/s-wave superconductor heterostructures have been predicted as candidate systems for the observation of Majorana fermions in the presence of superconducting vortices. In these systems, Majorana fermions are expected to form at the interface between the topological insulator and the superconductor while the bulk plays no role. Yet the bulk of a 3D topological insulator penetrated by a magnetic flux is not inert and can gap the surface vortex modes destroying their Majorana nature.  In this work, we demonstrate the circumstances under which only the surface physics is important and when the bulk physics plays an important role in the location and energy of the Majorana modes.
\end{abstract}
\maketitle
\section{Introduction}
Topological quantum computation is one of the most active areas of research in condensed matter physics. It promises to provide the advantages of quantum computation such as vast parallelism but with an inherent immunity from decoherence. This allows for the formation of qubits without the need for error correcting algorithms\cite{Kitaev20032}. The existence and stability of non-Abelian anyons forms the backbone of any architecture for topological quantum computation\cite{ivanov2001, nayak2008}. The simplest of these excitations is the Majorana fermion. Many diverse systems are predicted to harbor these heretofore elusive excitations including p-wave superconductors\cite{read2000,ivanov2001}, the $\nu=\frac{5}{2}$ fractional quantum Hall state\cite{Moore1991362}, and cold-atom systems\cite{tewari2007, regal2009}.

Recently, the search for Majorana fermions has expanded into the family of materials commonly referred to as topological insulators. Generally speaking, topological insulators are a class of materials with an insulating time-reversal invariant bandstructure for which strong spin orbit interactions lead to an inversion of the band gap at an odd number of time reversed points in the Brillouin zone. Topological insulators are differentiated from other ordinary band insulators by the presence of surface states containing Fermi arcs which encapsulate an odd number of Dirac points and are associated with a Berry's phase of $\pi$. Normally, such degeneracy points in the bandstructure are easily removed by any perturbations, but in the case of topological insulators the band crossing at the boundaries is protected because Kramer's theorem prevents time-reversal invariant perturbations from opening up a gap in the energy spectrum\cite{hasan2010}. In the inceptive work of Fu and Kane\cite{fu2008}, they show that coupling s-wave superconductors to 3D time-reversal invariant topological insulators\cite{fu2007b,moore2007,roy2009a,hasan2010} via the proximity effect may be a potential platform to realize these non-Abelian anyons. In particular, Fu and Kane show that the surface of a 3D topological insulator - s-wave superconductor heterostructure, exhibits many of the same properties as a chiral p-wave superconductor\cite{read2000} in that the cores of the vortex excitations may harbor Majorana fermions.

Nevertheless, while the analysis presented in Ref.\ \onlinecite{fu2008} considers the gapless surface-state Hamiltonian proximity coupled to a superconductor, it ignores the properties of the bulk topological insulator.  If the bulk were simply a trivial insulator it would be inert and no further considerations would be required. However, it is known that topological insulators react to the presence of thin flux tubes\cite{rosenberg2010,ostrovsky2010}, which can generate a ``worm-hole" effect that traps low-energy states on the flux tube. This is of particular concern as the simplest approach to create vortices in a 3D topological insulator - s-wave superconductor would be to coat the  surface of the 3D topological insulator with a type-II s-wave superconductor and then use magnetic flux tubes generated by an applied magnetic field to proliferate the vortices, which would then contain the Majorana states. In this work we seek to understand exactly when it is sufficient to only consider the surface physics, and when one must include the bulk physics. Very interesting work in this general direction is discussed in Ref.\ \onlinecite{Hosur:2011fk} where the role of chemical potential in the stability of the Majorana vortex modes is discussed for a topological insulator whose entire bulk has become superconducting. Here, instead, we focus only on the proximity effect scenario and the effects of the applied magnetic flux necessary to create a field of vortices.

The manuscript is organized in the following manner: In Section \ref{sec:model}, we detail the topological insulator Hamiltonian utilized in this work.  In Section \ref{sec:wormeffect}, we review the physics resulting from the addition of very thin magnetic flux lines in 3D topological insulators. In particular, we review how a magnetic flux line connect the surfaces of 3D topological insulators in which it enters and exits with a line of low-energy modes.   In Section \ref{sec:tisc}, we extend our analysis from the addition of magnetic flux lines in 3D topological insulators to topological insulators with s-wave superconducting pairing on the top and bottom surfaces. In this system, we discuss two different  physical regimes delineated by  the spread of the magnetic flux as it penetrates the heterostructure. In the first physical regime, we study the behavior of the topological insulator - superconductor heterostructure when the spread of the magnetic flux lines inserted into the system are limited in spatial extent to  a size on the order of the lattice constant.
This leads to the removal of the zero energy Majorana state from the system as the surface bound states may now tunnel along the magnetic flux tube and annihilate the states on the other surface.
In the second regime, we study the case when the spread of the magnetic flux line has a much wider spatial extent. In this situation, the Majorana fermions become localized at the interface between the topological insulator and proximity coupled superconductor and the bulk remains inert so that only the surface physics need be considered.

\section{Model Hamiltonian}
\label{sec:model}
 In order to capture the essential physics of the problem, we use a minimal bulk model for a 3D topological insulator which consists of a gapped Dirac Hamiltonian
\ba
H_{D}&=&\sum_{\pb}c^{\dagger}_{\pb}H_{D}(\pb)c^{\pdag}_{\pb}\nonumber\\
&=&\sum_{\pb}c^{\dagger}_{\pb}\left(d_{a}(\pb)\Gamma^{a}+M(\pb)\Gamma^0\right)c^{\pdag}_{\pb}.\label{eq:3dDirac}
\ea \noindent where $a=1,2,3,$ $\Gamma^{a}=\tau^{x}\otimes \sigma^{a},$ $\Gamma^{0}=\tau^{z}\otimes \mathbb{I},$ $\sigma^{a}$ is spin, $\tau^{a}$ is an orbital degree of freedom representing orbitals $A,B,$ and $c_{\pb}=(c_{\pb A\uparrow}\;c_{\pb A\downarrow}\;c_{\pb B\uparrow}\;c_{\pb B\downarrow})^{T}.$ In this work, to illustrate the salient physics, we will use both a continuum description with
\begin{equation}
d_{a}(\pb)=\hbar v_F p_a,M(\pb)=m-(1/2)bp^2
\end{equation}
and a lattice description with
\begin{equation}
d_{a}(\pb)=(\hbar v_F/a)\sin(p_a a),
\end{equation}
and
\begin{equation}
\begin{split}
M(\pb)=&(b/a^2)\left(\cos (p_x a)+\cos (p_y a)+\cos(p_z a)\right) \\
&-3b/a^2+m
\end{split}
\end{equation}
where $v_F,m,b$ are material parameters and $a$ is the lattice constant. These material parameters may be adjusted using the previously tabulated constants based on DFT calculations\cite{liu2010,Zhang:2009fk} to fit many of the most common 3D topological insulators. Here, to simplify the notation, we will set $a$ and $\hbar v_F$ equal to unity in the remainder of the work unless otherwise noted.  This model has time-reversal symmetry with $T=\mathbb{I}\otimes i\sigma^y K$ where $K$ is complex conjugation. For generic values of $m\neq 0$ the system is a gapped insulator and we focus on the low-energy regime when $m\sim 0.$. Assuming translation symmetry, the energy spectrum of the continuum model is $E_{\pm}=\pm\sqrt{ p^2+(m-(1/2)bp^2)^2}$ with each band doubly degenerate.  As a convention, which is consistent with the behavior in canonical topological insulators such as Bi$_{2}$Se$_{3}$, we choose $b>0$ and as a result the trivial (topological) insulator state occurs when $m/b<0$ ($m/b>0$). In the following, when we refer to a topological insulator state, we are referring to a state described by the model Eq. \ref{eq:3dDirac} with $m>0$ and $b>0.$

\section{Magnetic Flux Lines in 3D Topological Insulators}
\label{sec:wormeffect}
\begin{figure}
\includegraphics[width=8.0cm]{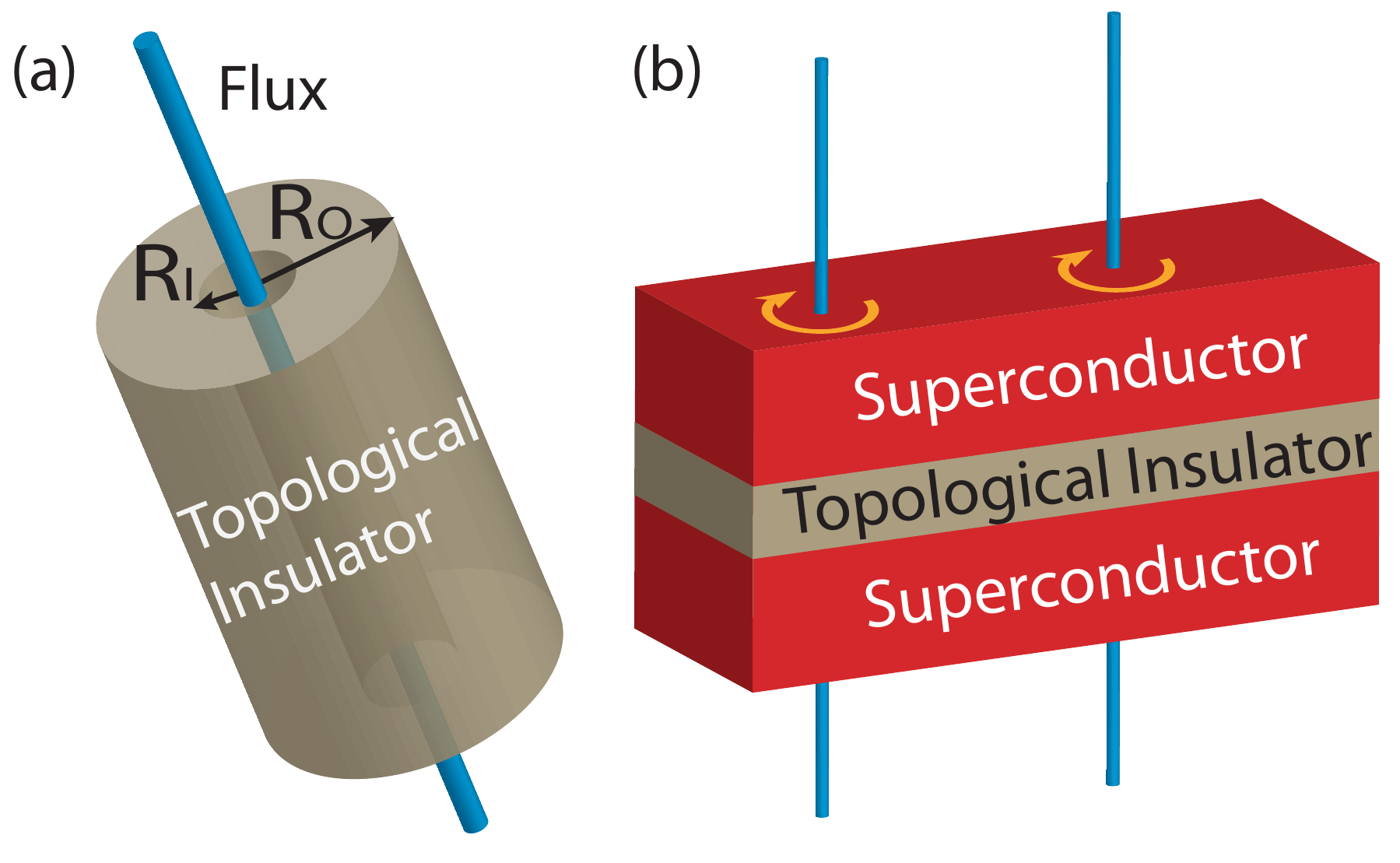}\\
\caption{(a) Schematic of a cylindrical 3D topological insulator with a hole drilled through the center. The blue line represents a flux tube threaded through the cylindrical hole.  (b) Schematic of a heterostructure of a topological insulator thin-film sandwhiched between two s-wave superconductors. The thin blue lines represent $h/2e$ flux tubes which generate vortices in the superconductor layers. }\label{fig:cylinder}
\end{figure}

The physics of thin flux lines in the bulk of a topological insulator was originally considered in Refs.\ \onlinecite{rosenberg2010,ostrovsky2010}. Let us begin with an infinite solid cylinder of 3D topological insulator whose length is placed along the z-direction with a cylindrical hole drilled through the center as seen in Fig.\ \ref{fig:cylinder}(a). We take the inner and outer radii to be $R_{I},R_{O}$ respectively.  Due to the characteristic property of time-reversal invariant topological insulators, there are low-energy modes bound to the inner and outer cylindrical surfaces. However, the surface fermions have a $\pi$-Berry phase when a particle winds around the Fermi surface. This leads to a condition that there will be no exact zero modes in the surface energy spectrum on the inner or outer surfaces, so long as we consider a cylinder of finite radius. To be clear, surface electrons that travel around the azimuthal direction on the inner or outer surfaces pick up a $\pi$-Berry phase leading to effective anti-periodic boundary conditions which shifts the zero-momentum Fourier mode away from zero energy. To recover the exact zero-modes we must twist the boundary conditions back to being periodic. This is accomplished by threading $\pi$-flux $(\phi_0/2=h/2e)$ through the hole drilled in the cylinder\cite{rosenberg2010,ostrovsky2010}.

In order to be concrete about the behavior of these zero energy modes, we begin with the continuum model Hamiltonian for a topological insulator introduced in Eq.\ \ref{eq:3dDirac} and assume the cylindrical with flux $\phi$ in the unit of $h/e$ threaded through the interior hole. Keeping only the linear terms in ${\textbf{p}}$ we get
\begin{align}
\label{eq:hline}
H=&m\tau_z\otimes \mathbb{I}+(p_x-eA_x)\tau_x\otimes\sigma_x \nonumber \\
&+(p_y-eA_y)\tau_x\otimes\sigma_y+p_z\tau_x\otimes\sigma_z.
\end{align}
In Eq.\ (\ref{eq:hline}), the vector potential is $\vec{A}=\frac{\phi}{ r^2}\frac{\hbar}{e}(-y\hat{x}+x\hat{y})$. As we have now added the magnetic flux into our Hamiltonian, momenta $p_x$ and $p_y$ are not no longer good quantum numbers. It is important to note that The Hamiltonian should be solved in real space where the momentum operators are represented as $p_x\ra-i\partial/\partial x$ and $p_y\ra-i\partial/\partial y$. First we will consider the case $p_z=0$ and solve for the zero-mode eigenstates. Converting from Cartesian coordinates $(x,y)$ to polar coordinates $(r,\theta)$ the Hamiltonian then becomes
\begin{eqnarray}
H_{linear}&=&
\begin{pmatrix}
m & 0   & 0 & P_{-\theta} \\
0  & m  & P_{\theta} & 0 \\
0  & P_{-\theta} & -m & 0 \\
P_{\theta} & 0  & 0 &-m
\end{pmatrix} \nonumber \\
P_\theta&=&e^{i\theta}\left[\frac{\partial}{i\partial r}+\frac{\partial}{r\partial \theta}-\frac{i\phi}{r}\right] \nonumber \\
\ P_{-\theta}&=&e^{-i\theta}\left[\frac{\partial}{i\partial r}-\frac{\partial}{r\partial \theta}+\frac{i\phi}{r}\right]. \label{Hlinear}
\end{eqnarray}
As we are searching for the zero energy modes in the system, we must solve the eigenvalue problem for this matrix. For the energies we obtain
\be
E_{\ell,\pm}=\pm \frac{\ell+\frac{1}{2}-\phi}{R_{I}}\label{eq:wormenergy}
\ee
with corresponding eigenstates
\begin{align}
\mid \psi_{E_+} \rangle=& \frac{ e^{-\int^r_{R_I} m(r')dr'}}{\alpha\sqrt{2r}}
(e^{i\ell\theta},0,0,ie^{i(\ell+1)\theta})^T, \nonumber \\
\mid \psi_{E_-} \rangle=& \frac{e^{-\int^r_{R_I} m(r')dr'}}{\alpha \sqrt{2r}}
(
0,-ie^{i(\ell+1)\theta},e^{i\ell\theta},0
)^T,
\end{align}
where $\alpha$ is a normalization coefficient defined as
\begin{equation}
\label{normconst}
\alpha^2=2\pi \int^\infty_{R_I}e^{-2\int^r_{R_I} m(r')dr'}dr.
\end{equation}
In Eq.\ \ref{eq:wormenergy} we note that $\ell$, an integer which represents an angular momentum quantum number though it should be noted that for the eigenstates, different components may possess different angular momenta. Thus, if the flux is $\phi=1/2+n$ for all integers $n$, then the resultant wavefunctions, $\psi_{0+}$ and $\psi_{0-}$ for $\ell=n$, have zero energy.


Following the procedure outlined in Ref. \onlinecite{rosenberg2010}, we now imagine adiabatically shrinking the radius $R_{I}\to a$ and consider a single lattice plaquette as the hole drilled through the center of the cylindrical topological insulator. Thus, $\pi$-flux threaded through a line of single-plaquettes produces zero modes localized on the line of plaquettes along the length of the cylinder on the inner and the outer boundary.  If we turn on $p_z$  we will find a Kramers' pair of propagating modes on the inner and outer surfaces with disperse linearly in $p_z.$ As we have time reversal invariance, we expect modes propagating in both directions with opposite spin polarizations. Therefore, a $\pi$-flux line confined to a hole, even in the limit where the hole is reduced to the size of a single plaquette, in a topological insulator will trap a single-pair of gapless counter propagating modes akin to the 1D holographic edge state found in a 2D quantum spin Hall system. This is referred to as the wormhole effect\cite{rosenberg2010} and it demonstrates that the bulk of a topological insulator is not generically inert when in the presence of magnetic flux.

\section{3D Topological Insulator - Superconductor Heterostructure}
\label{sec:tisc}

Having shown that the presence of magnetic flux in a 3D topological insulator forces one to consider the presence of a non-inert bulk, we proceed to understanding the effects of coupling type-II s-wave superconductors to the surfaces of a 3D topological insulator.  We will want to add a sufficient amount of magnetic flux to generate vortices yet not so much as to necessitate the consideration of the interactions or quasi-particle tunneling  between vortices. With proximity coupling to an s-wave superconductor, and in the presence of a magnetic field ${\textbf{B}},$ we must use the  Bogoliubov-de-Gennes (BdG) mean-field description of our Hamiltonian:
\begin{eqnarray}
H_{BdG}=\frac{1}{2}\sum_{\pb}\Psi^{\dagger}_{\pb}\left(\begin{array}{cc} H_{D}(\pb-e {\textbf{A}}) & \Delta \\ \Delta^{\dagger} & -H^{\ast}_{D}(-\pb-e{\textbf{A}})\end{array}\right)\Psi^{\pdag}_{\pb}\nonumber\\
\end{eqnarray}\noindent with $\nabla\times {\textbf{A}}={\textbf{B}},$  $\Delta=\Delta_0({\textbf{x}}) \mathbb{I}\otimes i\sigma^{y},$ and $\Psi_{\pb}=\left(c^{\pdag}_{\pb}\;\;\; c^{\dagger}_{-\pb}\right)^{T}.$  We consider a heterostructure geometry with a thin-film of topological insulator sandwiched along the z-direction between two s-wave superconductors as shown in Fig. \ref{fig:cylinder}(b). We model the physics of the  superconductors by inserting an induced s-wave pairing term into the BdG Hamiltonian which penetrates into the topological insulator film. Without the presence of a magnetic field, this implies that $\Delta_0 ({\textbf{x}})=\Delta_0(z)$. That is, we assume that that the superconducting proximity pairing is homogenous in the $xy$-plane if no vortices are present.

With this in mind, we can proceed with our analysis of the effects of the pairing. It is important to note that the pairing, if weaker than the energy scale of the bulk insulating gap, will not affect the gapped bulk states of the topological insulator. However, it will affect the metallic surface states which are susceptible to a superconducting pairing potential. The effective surface BdG Hamiltonian is
\begin{eqnarray}
H^{(surf)}_{BdG}=\frac{1}{2}\sum_{\pb}\Phi^{\dagger}_{\pb}\left(\begin{array}{cc} p_x\sigma^y-p_y\sigma^x & \Delta_0 i\sigma^y \\ -\Delta_{0}^{\ast}i\sigma^y & -p_x\sigma^y-p_y\sigma^x\end{array}\right)\Phi^{\pdag}_{\pb}\nonumber\\ \label{eq:surfbdg}
\end{eqnarray}\noindent where $\Phi_{\pb}=\left( c_{\pb\uparrow}\;\; c_{\pb\downarrow}\;\; c^{\dagger}_{-\pb\uparrow}\;\; c^{\dagger}_{-\pb\downarrow}\right)^{T}$ and $\Delta_{0}$ represents the effective pairing potential felt by the surface states. This Hamiltonian has a gapped energy spectrum $E_{\pm}=\pm\sqrt{p^2+\vert\Delta_0\vert^{2}}.$ Thus, a non-zero proximity coupling induces a gap in the topological surface states. Previous work has shown that a vortex induced on the proximity-coupled surface traps a Majorana bound state. This is shown by solving Eq. \ref{eq:surfbdg} with a vortex present, which is inserted by winding the superconducting order parameter $\Delta_0$ as
\begin{equation}
\Delta_0=\Delta_0(r)e^{i\theta(r)}
\end{equation}
where $\theta(r)$ is the polar angle\cite{fu2008}.

While it may be possible to find Majorana states at the center of vortices in 3D topological insulator - s-wave superconductor heterostructures, there is not a standard prescription of how to generate such vortices. We consider the simplest possible route and apply a uniform external magnetic field perpendicular to the heterostructure. Physically, we must apply a large enough magnetic field to generate vortices, but small enough that the vortex density is low as mentioned above.  As the topological insulator film is not inert to the addition of flux, we must be careful to account for the effects of the applied magnetic flux to ensure it does not spoil the bound state structure. For simplicity, we consider only the one (analytic results) and two-vortex (numerical results) problems noting that the one and two vortex problems are essentially equivalent, and only differ because of the choice of boundary conditions. If our system contains periodic boundary conditions in the x and y directions then we \emph{must} have an even number of vortices; this is the situation we consider in our numerics. If we choose open boundary conditions in x and y, then a single-vortex in the bulk implies the existence of another vortex at the boundary or at infinity; this is the case for our analytic results. We assume that a magnetic flux of only one $\phi_0$ quantum, parallel to the $z$-direction, penetrates the superconductors.  In our heterostructure the superconductors on top and bottom would, in principle, dynamically generate two vortices in each layer. We will assume that the induced vortices (that we put in by hand) are well-separated enough so that they do not influence each other and that the positions of the vortices on the top and bottom surfaces share the same $(x,y)$ position for simplicity.

Inside the superconductor the penetrating magnetic field satisfies the London equation\cite{deGennesBook}
\begin{equation}
{\textbf{B}}({\textbf{r}})-\lambda^2 \nabla^2 {\textbf{B}}({\textbf{r}})=\frac{\phi_0}{2}\delta({\textbf{r}})
\end{equation}\noindent near a vortex positioned at the origin with penetration depth $\lambda.$ The solution for ${\textbf{B}}({\textbf{r}})$ in the superconductor is
\begin{equation}
{\textbf{B}}({\textbf{r}})=\hat{z}\frac{\phi_0}{4\pi\lambda^2}K_0(r/\lambda)\label{eq:bofr}
\end{equation} where $K_n(x)$ are modified Bessel functions of the second-kind. The flux within the disk of the radius $r$ is
\be
\phi({\bf r})=\left(1/2-(r/2\lambda) K_1 (r/\lambda)\right).  \label{eq:flux}
\ee
 We want to model the effects of ${\textbf{B}}({\textbf{r}})$ in the entire heterostructure including the topological insulator but this not easy to account for. We instead opt for a more phenomenological approach to capture the qualitative physics. Once the flux leaves the superconducting layers and enters the topological insulator film it will spread out. For our purposes, we will consider a  model where $\lambda$  varies with the depth in the heterostructure as $\lambda=\lambda(z)$ and study how the vortex physics changes with $\lambda(z).$ If the film is thin, the flux will not have sufficient distance to spread before it must re-enter the top superconducting layer and thus modeling the insulator layer as having a finite penetration depth (larger than that of the superconductor) is not unreasonable. In order to understand the appropriate physics, it is natural, in the context of this problem to  consider two separate limits associated with the amount of magnetic flux penetration into the topological insulator. In the first limit, we wish to examine the ``thin-flux" limit in which the flux which penetrates the topological insulator does not spread out very far in the topological insulator before reentering the other superconducting layer. In the second limit, we examine the case in which the magnetic flux spreads out widely in the topological insulator before it mush re-enter the other superconducting layer. These two limits can be considered analytically while we provide numerical calculations which capture the interpolation between these cases.  

\subsection{Thin-Flux Limit ($\lambda\sim a$)}

We begin from the limit of two well-separated, thin flux tubes of flux $\phi_0/2$ where the flux tubes are each confined within single plaquettes \emph{i.e.} $\lambda\ll a.$ When the proximity pairing potential vanishes, the system will exhibit gapless modes propagating on each of the thin flux tubes (ignoring finite size splitting due to hybridization with the second vortex). The resulting gapless theory of a single tube is simple to understand using the results from the previous section. There we solved Eq.\ \ref{eq:3dDirac} at $p_z=0$ with a $\pi$-flux tube through a single plaquette to obtain the two zero-mode solutions $\psi_{0+},\psi_{0-}$ (a Kramers' pair) localized on the flux. Then we can use ${\textbf{k}}\cdot {\textbf{P}}$ perturbation theory and treat $p_z$ as a perturbation to obtain, in the basis of $\psi_{0+},\psi_{0-}$, the low-energy Hamiltonian $H_{flux-line}=p_z\sigma^x$. This Hamiltonian is identical to the edge Hamiltonian of a quantum spin Hall edge state, as mentioned earlier. If we begin to increase $\lambda$ which corresponds to allowing the magnetic flux to spread uniformly in the z-direction, \emph{i.e.}\ we move away from the wormhole limit, this applies a perturbation to the gapless flux-line Hamiltonian. Using perturbation theory, we find that
\begin{equation}
H_{flux-line}=p_z\sigma^x+m_x(\lambda)\sigma^{y}+m_y(\lambda)\sigma^{z} \label{fluxline}
\end{equation}\noindent where the mass term, $m_{i}$, is monotonically increasing as $\lambda$ increases. This Hamiltonian has a gapped energy spectrum $E_{\pm}=\pm\sqrt{p_{z}^2+m_{x}^2+m_{y}^2}$ which is expected since time-reversal is broken and the flux-line Kramers' degeneracy at $p_z=0$ is lifted. Note that we are not increasing the \emph{amount} of flux, only the region over which it spreads. If the $\pi$-flux tube is larger than one plaquette then some bonds in the lattice model will necessarily have phase factors which have imaginary contributions to the Hamiltonian, regardless of the gauge choice, which break the time-reversal symmetry of the system. Our perturbation theory analysis is approximately valid until the induced gap $E_{M}=\sqrt{m_{x}^2+m_{y}^2}$ approaches the bulk mass gap $m.$ To estimate the size of the induced gap, $E_M$, caused by the spreading of the flux in the topological insulator ($\phi({\bf r})$) in Eq.\ \ref{eq:flux}, we apply first-order perturbation theory:
\be
E_M=\langle \psi_{0+}\mid\Delta H \mid \psi_{0+}\rangle=-\langle \psi_{0-}\mid\Delta H \mid \psi_{0-}\rangle,
\ee
where $\Delta H=H_{linear}(\phi({\bf r}))-H_{linear}(\phi=1/2)$. Using the previously obtained expression for $H_{linear}$ in Eq.\ \ref{Hlinear}, the first order approximation for $E_M$ in the continuum limit is
\be
E_M= \frac{\pi}{\lambda \alpha^2}\int_a^{\infty}K_1(r/\lambda)e^{-2\int^r_a m dr'}dr
\ee

We refer to this flux regime as the `thin-flux limit' \emph{i.e.}\ the regime in which we can consider the low-energy states as those originating from the gapped wormhole modes.
\begin{figure*}
\includegraphics[width=18cm, height=6.83cm]{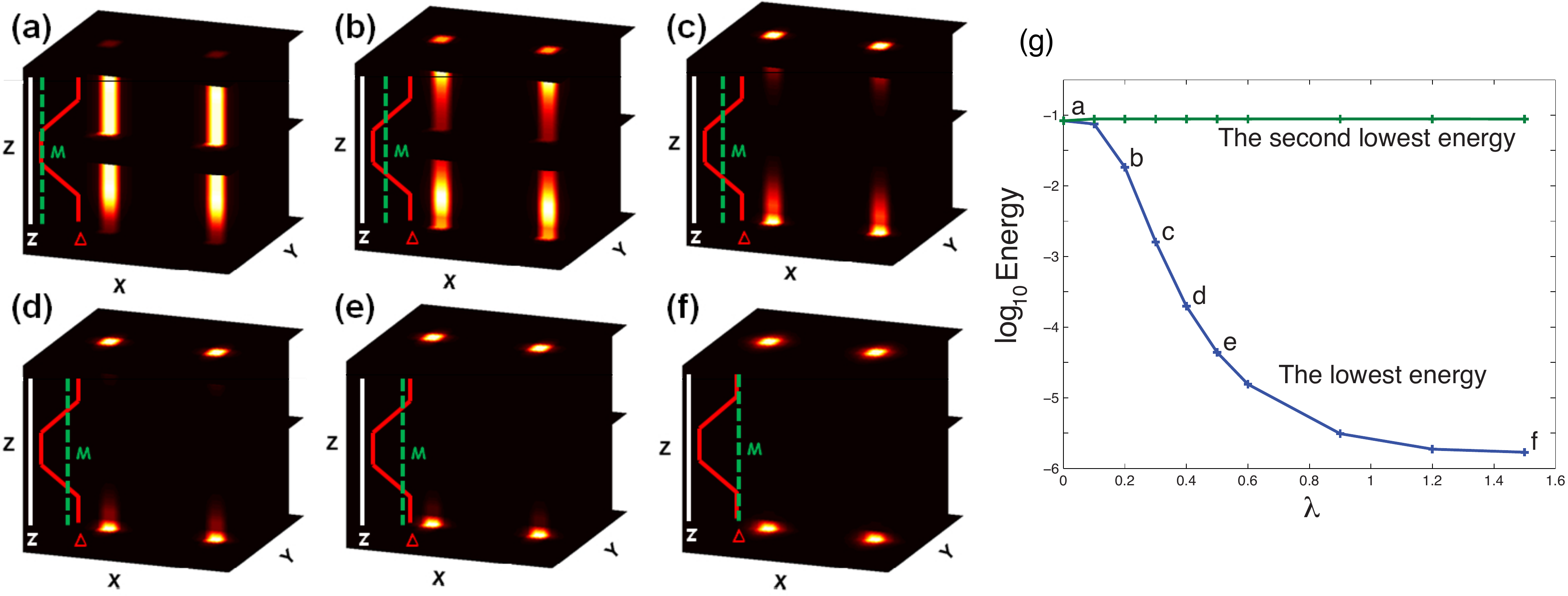}\\
\caption{Probability distribution for the lowest energy states corresponding to different superconducting penetration depths, $\lambda$ with: (a) $\lambda = 0.001$, (b) $\lambda = 0.2$, (c) $\lambda = 0.3$, (d) $\lambda = 0.4$, (e) $\lambda = 0.5$, (f) $\lambda = 1.5$.  Note that $\lambda$ is in units of the lattice constant $a$ and the entire flux is spread out in a region with a radius of roughly  $~5\lambda.$ The pairing potential $\Delta(z)$ decays from $0.5$ 0n the surface to $0$ within 5 layers. As $\lambda$ is increased we see that the states move from being delocalized along the flux tube penetrating the bulk of the topological insulator to being pinned at the surface. The inset shows a schematic of the spatial variation of the superconducting mass and the time-reversal breaking mass associated with the magnetization as $\lambda$ varies. (g)The energies of the lowest energy states as a function of $\lambda.$ There are clear zero modes forming as $\lambda\to\infty.$ Our numerical results show that in $\lambda\ll a$ regime, the lowest energy linearly decays as the height of the sample increases; in $\lambda\gg a$ regime, the lowest energy exponentially decays as the height of the sample increases. }\label{fig:domainwallschematic}
\end{figure*}
With the effects of the magnetic flux accounted for, we now turn on the superconductor proximity effect in the thin-flux limit. There will be an induced superconducting pairing potential that is $z$-dependent and, when flux and the corresponding vortices are present, the superconducting pairing takes on an $x$ and $y$ dependence as well. Before we get to the situation where $\Delta_0$ is only non-vanishing near the top and bottom surfaces, let us consider an induced $\Delta_0$ which is homogenous in the $z$-direction over the entire topological insulator. In the thin-flux limit the only low-energy metallic degrees of freedom are localized near the flux line so we can use our effective flux-line Hamiltonian from Eq. \ref{fluxline} to form a BdG Hamiltonian for the low-energy degrees of freedom:
\begin{equation}
\label{eq:hbdgpair}
H^{(BdG)}_{flux-line}(p)=\frac{1}{2}\left(\begin{array}{cc}H_{flux-line}(p)& i\Delta_{0}\sigma^{y}\\ -i\Delta_{0}^{\ast}\sigma^y & -H^{\ast}_{flux-line}(-p)\end{array}\right)\nonumber
\end{equation}\noindent which has an energy spectrum with four non-degenerate bands 
\begin{equation}
\pm E_{\pm}=\pm\sqrt{p_{z}^2+ \left(\vert\Delta_0\vert\pm E_{M}\right)^2}.
\end{equation}
This spectrum is gapped unless $\vert\Delta_0\vert=\vert E_{M}\vert.$ Now let us consider more realistic conditions where the thin-film is too thick to become entirely superconducting, and the proximity induced pairing depends on $z$. Specifically, the superconducting pairing potential decays as we move away from the surfaces towards the interior the topological insulator film. In this case the system has a time-reversal symmetry breaking mass $E_M$ which is homogenous in the $z$-direction (as per our phenomenological model) and a superconducting mass, $\vert\Delta_0(z)\vert$, which is $z$-dependent. From standard 1D Dirac physics\cite{Jackiw1976,fu2008} this model will exhibit localized, zero-energy  Majorana bound states on mass domain walls in the z-direction along the flux line \emph{i.e.} the places where $\vert\Delta_0 (z)\vert=\vert E_{M}\vert.$ As the thickness of the flux increases so does $E_{M}$ and the domain-walls along the vortex line get pushed toward the surface.

This perturbation theory is valid as long as $E_{M}\ll m.$  If we want to be able to carry out a full interpolation  between the thin-flux limit and the thick flux limit (to be discussed in the next section) we must rely on a numerical calculation. We show the results of such a calculation in Fig. \ref{fig:domainwallschematic}. We used a Lanczos exact-diagonalization algorithm to solve for the zero-modes of a full 3D lattice model. The vortices and proximity effect associated with the superconducting regions were non-dynamical and included in the mean-field limit following, for example, Ref. \onlinecite{vafek2001}. We present the details of our numerical calculations in Appendix A. As illustrated in Fig. \ref{fig:domainwallschematic}a-f, when we allow the flux to spread in the topological insulator film, \emph{i.e.} as $\lambda$ increases, the domain-wall bound states, which begin in the interior of the topological insulator, move outward toward the surface. As discussed in the previous paragraph this can be understood by noting that as $\lambda$ increases $E_M$ increases and the position of the mass-domain wall moves toward the surface. As the flux becomes thicker the bound states become more localized on the surfaces at the points around which the superconducting order parameter winds due to the flux. Since there are two domain-walls on each flux tube the pair of bound states will hybridize and lie higher than zero energy as shown in Fig. \ref{fig:domainwallschematic}g. As $\lambda$ increases the hybridization decreases which rapidly drives the states towards zero energy.

\subsection{Thick-Flux Limit ($\lambda\gg a$)}

From Fig.\ref{fig:domainwallschematic}, we see that in the extreme thin-flux limit, the Majorana modes will penetrate into the bulk and hybridize with the states on the other surface and annihilate. Fortunately, as indicated by our analytic perturbation theory, and numeric lattice model calculations, $\lambda$ does not have to be very large before we move from the wormhole effect/thin-flux limit so that the vortex modes are tightly bound to the surface at zero energy. The wormhole effect simply generates a region in the bulk with a mini-gap across which the Majorana states can tunnel to the opposite surface. While the bulk of a topological insulator is not inert to flux insertion, as long as the flux is not tightly bound to a region on the order of a lattice plaquette, the Majorana states will have difficulty tunneling between the top and bottom surfaces and will be well-localized in the surface vortex cores.

Once the flux is thick enough to restore the bulk gap entirely, we can consider the explicit Majorana bound state solution in the presence of the magnetic flux from the surface Hamiltonian alone.  We consider the BdG surface-state Hamiltonian in Eq. \ref{eq:surfbdg} with non-zero vortex winding and magnetic flux. We focus on the neighborhood of a single vortex and solve the problem for generic flux and order parameter profiles in the continuum limit. We begin by assuming we have a vortex at the origin generated by a magnetic flux given by Eq. \ref{eq:bofr}. The surface Dirac Hamiltonian is
\begin{eqnarray}
H&=&\left(\begin{array}{cc}
H(p,\phi) & i\sigma^y \Delta_0 e^{-i\theta}\\
-i\sigma^y \Delta^{\ast}_0 e^{i\theta} &-H^{\ast}(-p,\phi)\end{array}\right)\\
H(p,\phi)&=&\nonumber\\
&\hbar v_{F}&\left(\begin{array}{cc} 0 & -e^{-i\theta}\left(\partial_r+\frac{-i\partial_{\theta}+\phi}{ r}\right)\\
e^{i\theta}\left(\partial_r-\frac{-i\partial_{\theta}+\phi}{ r}\right)&0 \end{array}\right)\nonumber \label{surfH}
\end{eqnarray}
\noindent where we have changed to polar coordinates and have implemented a non-zero vector potential $A_{\theta}=\hbar \phi({\bf r})/e r$ where $\phi (\textbf{r})$ is given in Eq. \ref{eq:flux}. This Hamiltonian has \emph{two} eigenstates with zero energy:
\begin{eqnarray}
\vert \psi_1 \rangle& =&
\frac{1}{\beta} e^{-\int_0^r\left(\frac{\Delta(r')}{\hbar v_{F}}+\frac{\phi(r')}{r'}\right) dr'}\left( \begin{array}{c}
0  \\
1  \\
0 \\
1
\end{array} \right), \\
\vert \psi_2 \rangle &=&\frac{1}{\beta} e^{-\int_0^r\left(\frac{\Delta(r')}{\hbar v_{F}}+\frac{1-\phi(r')}{r'}\right) dr'}\left( \begin{array}{c}
e^{-i\theta}  \\
0 \\
e^{i\theta} \\
0
\end{array} \right),
\end{eqnarray}
where $\beta$ is a normalization coefficient.  If we turn off the magnetic field \emph{i.e.}\  for $\phi=0$ we are in the Fu-Kane limit with very diffuse flux and then only the state $\vert\psi_1\rangle$ is normalizable which matches their result\cite{fu2008}. For a generic flux profile there will still only be one zero-mode solution that satisfies the boundary conditions and it will be a linear combination $\vert\psi_0\rangle=a_{1}(\phi)\vert\psi_1\rangle + a_2(\phi)\vert\psi_2\rangle.$ The coefficients $a_1(\phi),a_2(\phi)$ control the spin composition of the zero-mode and depend on not only the detailed boundary conditions but also on the short-distance physics of the vortex structure.

\begin{figure}[htp]
  \begin{center}
	\includegraphics[width=70mm]{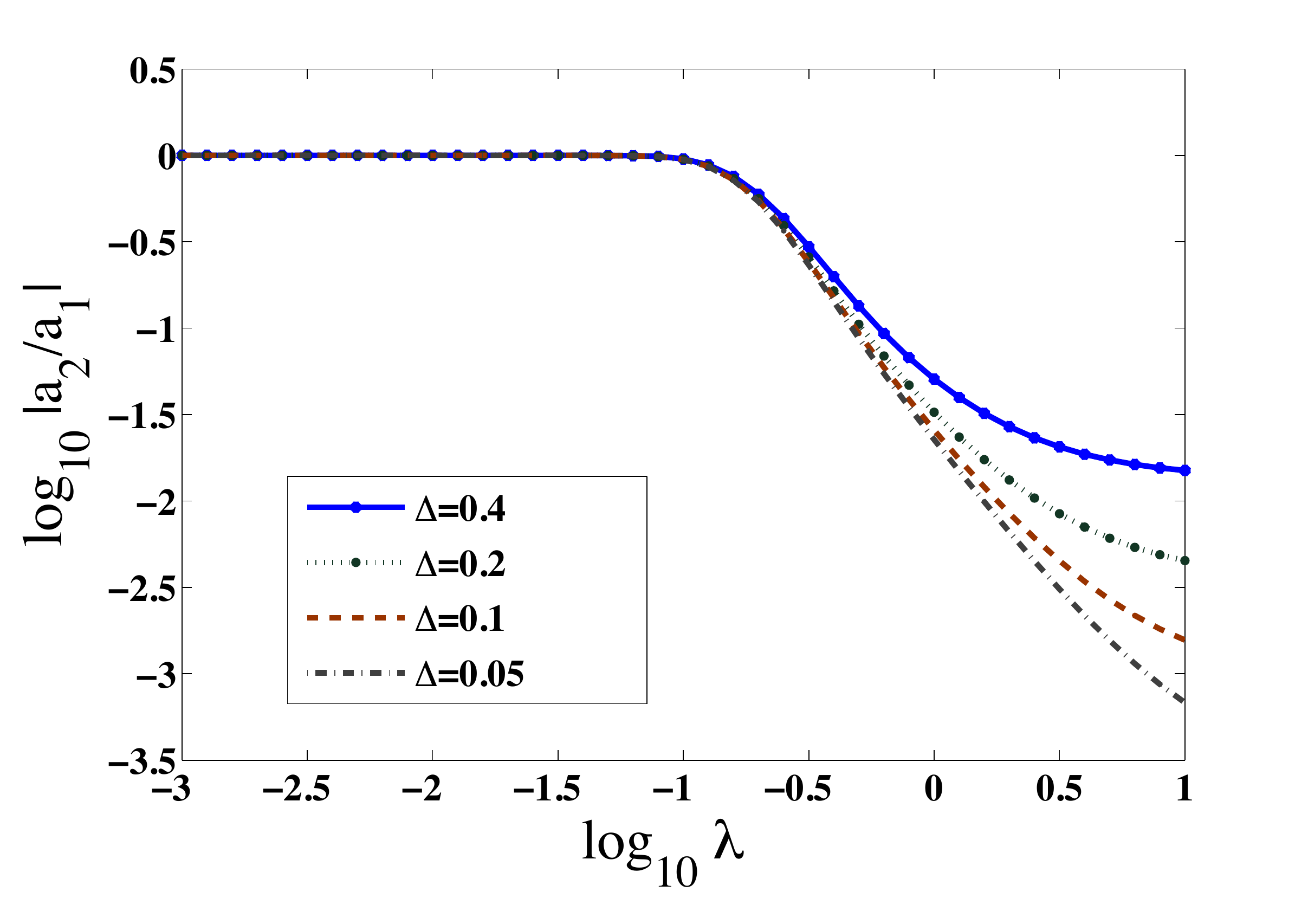}
  \end{center}
  \caption{ Ratio of spin up to spin down composition of a Majorana bound state versus flux penetration depth $\lambda.$ The different traces represent the use of different superconducting pairing potential strengths ($\Delta$). In the thick-flux limit, a larger $\Delta$ corresponds smaller ratio of spin up and down leading to spin polarized Majorana bound states.}
  \label{lambda}
\end{figure}

Since the coefficients $a_1(\phi), a_2(\phi)$ depend on the details of the system we numerically calculate them.  By solving the eigenvalue problem of $H_{BdG}^{surf}$ from Eq. \ref{eq:hbdgpair}, a zero mode Majorana bound state exists in the core of a vortex.  The ratio of spin up to spin down ($|a_2|/|a_1|$) for a single zero mode is shown in Fig.\ \ref{lambda}. This ratio describes the mixing between the two allowed zero energy modes when finite flux is present. We find that at $\lambda\sim 0.1$, where the magnetic flux starts to spread over more that one plaquette, the ratio of spin up and down starts to decrease rapidly. As $\lambda\rightarrow \infty$, $|a_2|/|a_1|\rightarrow 0$. In this limit, this is a Fu-Kane Majorana bound state \cite{Fu:2008fk} possessing a single species of spin. As $\lambda \rightarrow 0$, $|a_2|/|a_1|\rightarrow 1.$ This limit is the thin-flux limit, for which we see that the zero modes have the same portions of spin up and down. This is due to the fact that in Eq.\ \ref{fluxline} the Hamiltonian $H_{flux-line}$ contains equal portions of spin up and down, and, therefore, the zero modes in this limit have the same portions of spin up and down. In short, as $\lambda$ decreases we move from the thick-flux to thin-flux limits and $|a_1(\phi)| (|a_2(\phi)|)$ monotonically decreases (increases). Thus, we find that in the limit where the flux does not affect the bulk physics the effective magnetic field only acts to change the spin composition of the zero-energy vortex core state.

Beyond understanding the spatial extent of the flux spread on the Majorana states, we wish to look at changing $\Delta$, which changes the extent of the Majorana bound state. We fix the penetration depth of the flux to lie in the thick-flux regime and consider the probability distribution of the zero modes. In the thick flux limit, we find that the spin down dominates the composition of the Majorana bound states. Therefore, zero-mode wavefunction, described in Eq.\ \ref{surfH}, has a probability distribution
\begin{equation}
P(r)=\langle \psi_1 \mid \psi_1 \rangle=\frac{2}{r \beta^2 }e^{-2\Delta r/\hbar v_F - K_0(r/\lambda)}. \label{prodist}
\end{equation}
The decay length of the probability distribution approximately equals $\hbar v_F/\Delta$ for $\lambda\gg\hbar v_F/\Delta.$  The reason is that as $x\rightarrow 0$, $K_0(x)\sim -\ln x$ so then $P(r)\sim e^{-2\Delta r/\hbar v_F}$ which is the Fu-Kane result\cite{fu2008}. However, for $\lambda < \hbar v_F/\Delta$, we have to consider the probability distribution in Eq.\ \ref{prodist} directly to find the width of the zero modes.
\begin{figure}
  \begin{center}
	\includegraphics[width=70mm]{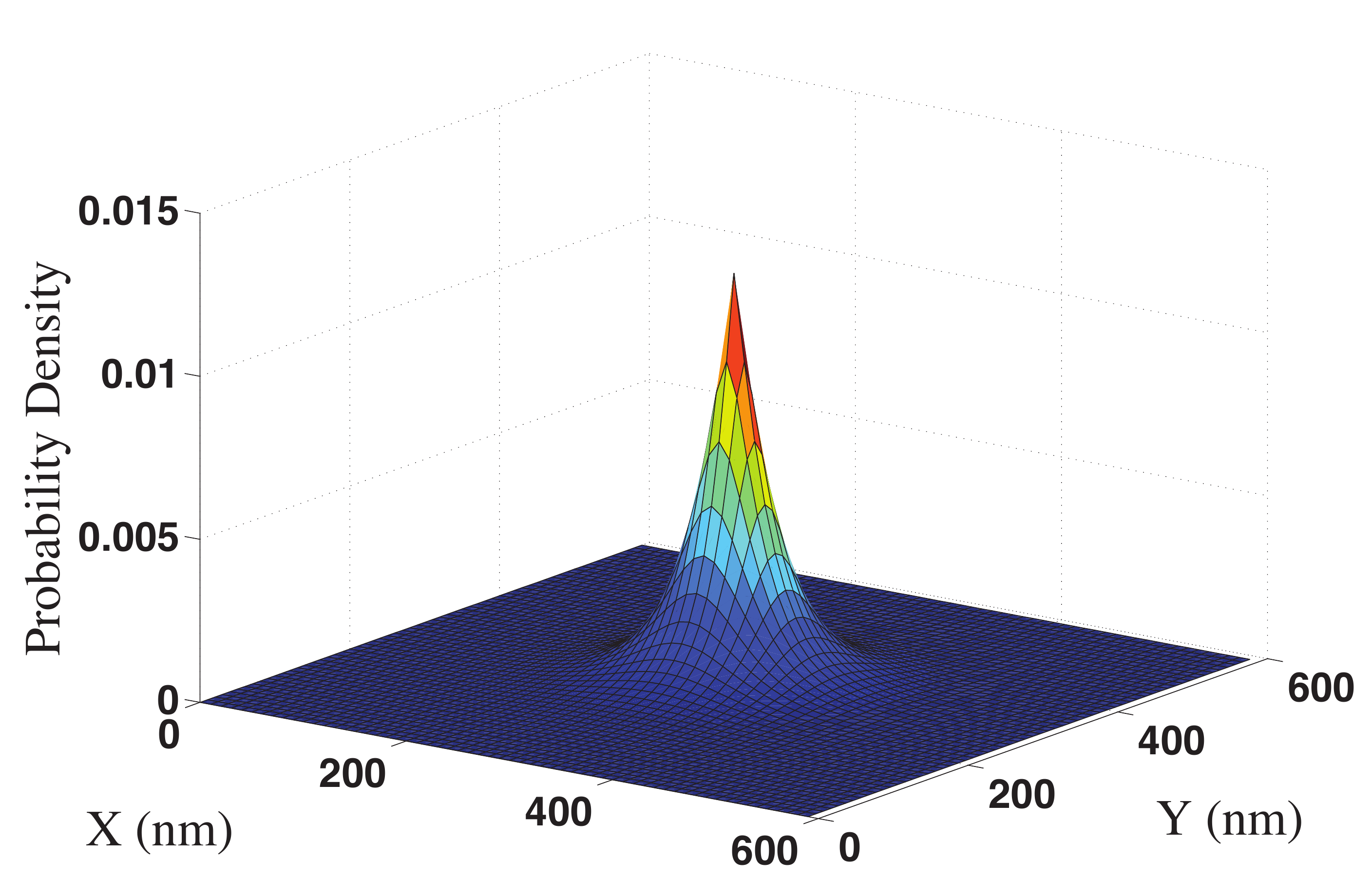}
  \end{center}
  \caption{Probability distribution of a Majorana bound state in the in the topological insulator / s-wave superconductor heterostructure where we have used realistic materials parameter. For the topological insulator we have used the  parameters of $\text{Bi}_2\text{Se}_3$ and for the superconducting films on the top and bottom surfaces we have used the material parameters of a niobium superconductor with a single vortex. }
  \label{realcase}
\end{figure}

It is important to solidify these results by discussing our results within a realistic context. Therefore, we estimate the width of Majorana bound states with real physical parameters. According to Ref.\ \onlinecite{Zhang:2009fk}, for the $\text{Bi}_2\text{Se}_3$ family of topological insulators $\hbar v_F \sim 4 eV \AA$.  For the superconducting top and bottom layers we use the pairing potential of the type-II superconductor niobium which approximately equals 1 meV. We assume that the proximity effect, which results in a pairing potential on the surface of the topological insulator has an induced pairing also of the order of $1$ meV which is the best-case scenario. Using these parameters, the quantity $\hbar v_F/\Delta$ is about $400$ nm, which is much larger than the London penetration depth of $40$ nm \cite{Maxfield:1965uq}. This combination of parameters allows us to now plot the probability distribution from Eq.\ \ref{prodist} directly as shown in Fig.\ \ref{realcase}. We find that in this case, the width of a Majorana bound state is of the same order of magnitude as that of the penetration depth. In general, the Majorana bound states will have the same characteristics as shown in Fig.\ \ref{realcase} for $\lambda < \hbar v_F/\Delta$. In most experimental regimes the physics will be deep in the thick-flux region where only the surface physics is important. One notable exception would be experiments where vortex pinning sites are artificially created (\emph{e.g.} by drilling through the heterostructure). In this case there is a possibility that the holes inside the topological insulator could trap a sizable fraction of a $\pi$-flux quantum which will lead to a mini-gap region across which the majorana bound states can tunnel. 

\section{Conclusion}

In conclusion, we have illustrated the effects of magnetic flux in topological insulator/superconductor heterostructures in two different regimes. In the thin-flux limit the Majorana fermion bound states can be destabilized through hybridization with low-energy bulk states localized near the thin flux line. These effects will be more pronounced in thin topological insulator films with minimal flux spreading, or in samples where vortex pinning sites are produced by drilling holes through the heterostructures. If such holes continue to trap approximately $h/2e$ flux throughout the topological insulator film then effects of flux in the bulk must be carefully considered. The opposite regime, where the vortex core does not feel much effective flux is likely the physical regime of most experiments. In this case we can ignore the bulk effects and focus only on the surface. The flux in this regime simply acts to change the spin content of the vortex zero mode and does not affect its energy or stability.  Furthermore, when we consider the parameters corresponding to real topological insulator / s-wave superconductor heterostructure with  $\text{Bi}_2\text{Se}_3$ as the topological insulator and with niobium superconductor layers, we find that a Majorana fermion trapped in the magnetic flux is stable with a spatial extent of around $40$ nm.

\emph{Acknowledgements}
We acknowledge useful conversations with B. Andrei Bernevig. C-KC is supported by the NSF under grant DMR 09-0329. MJG is supported by the AFOSR under grant FA9550-10-1-0459. TLH is supported by the NSF, under grant DMR 0758462 at the University of Illinois, and by the ICMT.


\appendix
\section{Numerical Calculations for the Bulk Hamiltonian}
We construct the Hamiltonian of a strong topological insulator sandwiched between two s-wave superconductors with vortices. The lattice Dirac model we use for the 3D topological insulator is
\begin{align}
H_D=\sum_{\bf r} H(m)c^\dagger_{\bf r}c_{\bf r}+\sum_{{\bf r},\delta}H(\delta)c^\dagger_{\bf r}c_{{\bf r}+\delta} \\
H(m)=m-3b, \ \ H(\delta)=\frac{b\Gamma_0+iA\hat{\delta}\cdot\vec{\Gamma}}{2}, \nonumber
\end{align}
where $r$ indicates the position of the lattice, $\delta(=\pm a \hat{x},\ \pm a \hat{y},$ $\pm a \hat{z}$) indicates the nearest neighbor hopping, and $\vec{\Gamma}=\Gamma_1\hat{x}+\Gamma_2\hat{y}+\Gamma_3\hat{z}$. Consider the interface between a strong topological insulator and an s-wave type II superconductor with an even number of vortices. The proximity effect leads to the pairing potential of the superconductor leaking into the topological insulator. Near the interface of the topological insulator we can write down the $8\times 8$ BCS-type lattice BdG Hamiltonian
\begin{widetext}
\begin{align}
H_{BdG}&=
\sum_{r}
\left( \begin{array}{cc}
c_{{\bf r}}^\dagger & c_{{\bf r}}
\end{array} \right)
\left( \begin{array}{cc}
H(m) & \Delta_0\mathbb{I}\otimes i\sigma_y e^{i\phi({\bf r})} \\
-\Delta_0\mathbb{I}\otimes i\sigma_ye^{-i\phi({\bf r})} & -H^*(m)
\end{array} \right)
\left( \begin{array}{c}
c_{{\bf r}} \\
c_{{\bf r}}^\dagger
\end{array} \right)  \nonumber \\
&+
\sum_{r,\delta}\left( \begin{array}{cc}
c_{{\bf r}}^\dagger & c_{{\bf r}}
\end{array} \right)
\left( \begin{array}{cc}
H(\delta)e^{- i\frac{e}{\hbar }\int^{{\bf r}+\delta}_{{\bf r}} {\bf A}({\bf r})\cdot d{\bf l}} & 0  \\
0 & -H^*(\delta)e^{i\frac{e}{\hbar }\int^{{\bf r}+\delta}_{{\bf r}} {\bf A}({\bf r})\cdot d{\bf l}}
\end{array} \right)
\left( \begin{array}{c}
c_{{\bf r}+\delta} \\
c_{{\bf r}+\delta}^\dagger
\end{array} \right), \label{Hnumeric}
\end{align}
where ${\bf A}$ is the vector potential is coming from the magnetic field $B$ described by the conventional London equation with penetration depth $\lambda,$ as in Eq.\ \ref{eq:bofr}, with the cores of the vortices at $r_j$. The phase $\phi({\bf r})$ acts as an additional ``gauge field'' coupled to the quasiparticles. With vortices, $\phi({\bf r})$ is not a pure gauge: $\nabla\times \nabla \phi({\bf r})=2 \pi \hat{z}\sum_j\delta({\bf r}-r_j)$. For the numerical calculation, we want to cancel out the phase $\phi({\bf r})$ in the order parameter to speed-up the simulation. Although $\phi({\bf r})$ is not a pure gauge, by performing a ``bipartite'' singular gauge transformation \cite{PhysRevLett.84.554}, the phase is successfully moved to the diagonal terms of the Hamiltonian. Here, in our main numerical calculation, we consider two vortices located at $\rb^A$ and $\rb^B$ respectively. A ``bipartite'' singular gauge transformation is $c_{r}=c'_{r}e^{i\phi_A({\bf r}_i)}$ for the particle part and $c_{r}=c'_{r}e^{i\phi_B({\bf r}_i)}$ for the hole part, where $\nabla\times \nabla \phi_A({\bf r})=2\pi \hat{z}\delta({\bf r}-{\bf r}^A)$ and $\nabla\times \nabla \phi_B({\bf r})=2\pi \hat{z}\delta({\bf r}-{\bf r}^B)$. This gauge transformation avoids a multi-valued problem so that the integral $\int_{{\bf r}}^{{\bf r}+\delta}\nabla\phi_{A/B}({\bf r})\cdot d{\bf l}=\phi_{A/B}({\bf r}+\delta)-\phi_{A/B}({\bf r})$ is path-independent up to $2\pi n$, which does not affect the probability distributions of the eigenstates from the numerical calculation. The Hamiltonian we use in the simulation becomes
\begin{align}
H_{BdG}&=
\sum_{{\bf r}}
\left( \begin{array}{cc}
c_{{\bf r}}^\dagger & c_{{\bf r}}
\end{array} \right)
\left( \begin{array}{cc}
H(m) & \Delta_0\mathbb{I}\otimes i\sigma_y \\
-\Delta_0\mathbb{I}\otimes i\sigma_y & -H^*(m)
\end{array} \right)
\left( \begin{array}{c}
c_{{\bf r}} \\
c_{{\bf r}}^\dagger
\end{array} \right)  \nonumber \\
&+
\sum_{r,\delta}\left( \begin{array}{cc}
c_{{\bf r}}^\dagger & c_{{\bf r}}
\end{array} \right)
\left( \begin{array}{cc}
H(\delta)e^{i\int^{{\bf r}+\delta}_{{\bf r}} [\nabla \phi_A ({\bf r})-\frac{e}{\hbar c}{\bf A}({\bf r})]\cdot d{\bf l}} & 0  \\
0 & -H^*(\delta)e^{-i\int^{{\bf r} +\delta}_{{\bf r}}  [\nabla \phi_B ({\bf r})-\frac{e}{\hbar c}{\bf A}(r)]\cdot d{\bf l}}
\end{array} \right)
\left( \begin{array}{c}
c_{{\bf r}+\delta} \\
c_{{\bf r}+\delta}^\dagger
\end{array} \right), \label{Hnumeric}
\end{align}
where
\begin{align}
&\int^{{\bf r} +\delta}_{{\bf r}}  [\nabla \phi_{A/B} ({\bf r})-\frac{e}{\hbar }{\bf A}({\bf r})]\cdot d{\bf l}= \nonumber \\
&\frac{1}{2}\int_\rb^{\rb+\delta}[(-\frac{(y-y^{A/B})}{|\rb-\rb^{A/B}|^2}\hat{x}+\frac{(x-x^{A/B})}{|\rb-\rb^{A/B}|^2}\hat{y})-(-\frac{(y-y^{B/A})}{|\rb-\rb^{B/A}|^2}\hat{x}+\frac{(x-x^{B/A})}{|\rb-\rb^{B/A}|^2}\hat{y}) \nonumber \\
&+\frac{1}{\lambda}\sum_{\nu=A,B}(-\frac{(y-y^\nu)}{|\rb-\rb^\nu |}\hat{x}+\frac{(x-x^\nu)}{|\rb-\rb^\nu |}\hat{y})K_1(\frac{|\rb-\rb^\nu|}{\lambda})]\cdot d{\bf l}.
\end{align}
\end{widetext}
The size of the topological insulator is $(n_x-1)\times (n_y-1) \times (n_z-1)$. In the numerical calculation, we typically used  $n_x=28$, $n_y=20$, and $n_z=24$ with the lattice constant $a=1$, also with open boundary conditions in all directions. We are modeling a thin-film of topological insulator  sandwiched along the z-direction between two s-wave superconductors. Because of the proximity effect, we assume $\Delta_0=\Delta_0(z)$ smoothly decays away from the top and the bottom surfaces and vanishes in the middle region. In the xy-plane, let the center be the origin (0,0). The positions of the two vortices are set at $(nx/4,0)$ and $(-nx/4,0)$. Also in the thin film limit, we assume phenomenologically that the penetration depth $\lambda$ is independent of $z$. We choose $m=1.5$, $b=1$, and $A=\hbar v_F=1$. This is a strong topological insulator phase with topological invariants $(1;111)$. Finally, after solving the eigenvalue problem to find the lowest energy modes of the Hamiltonian in Eq.\ \ref{Hnumeric}, the probability distributions of the lowest energy modes with varying $\lambda$ are shown in Fig.\ \ref{fig:domainwallschematic}.

\section{Numerical Calculations for the Surface Hamiltonian}
In the thick-flux limit, the zero modes are pushed to the surface of the topological insulator. Therefore, the surface  Hamiltonian can adequately describe the physics of the zero modes. Solving an eigenvalue problem of the 2D Hamiltonian allows us to consider a larger size of the system. In the following, we will derive the surface Hamiltonian from the bulk BdG Hamiltonian in Eq.\ \ref{Hnumeric} with vanishing magnetic field, and transform it to position space for the simulation. The bulk BdG Hamiltonian can be written explicitly
\begin{align}
&H_{BdG}= \nonumber \\
&\frac{1}{2}(M({\bf p})\sigma_z\otimes \Gamma^0+\sin p_x\mathbb{I}\otimes \Gamma^1+\sin p_y\sigma_z\otimes \Gamma^2 \nonumber \\
&+\sin p_z \mathbb{I}\otimes \Gamma^3+\Delta_R\sigma_x\otimes \mathbb{I}\otimes i\sigma_y-\Delta_I\sigma_y\otimes \mathbb{I}\otimes i\sigma_y),
\end{align}
where $\Delta_R$ and $\Delta_I$ are the real and imaginary parts of the order parameter. We start by finding the zero modes on the surface of a strong topological insulator. Therefore, to find the surface/domain-wall zero modes we need to have $H_z\vert\psi\rangle=(M({\bf p})\sigma_z\otimes \Gamma^0+\sin p_z \mathbb{I}\otimes \Gamma^3)\vert\psi\rangle=0.$  Qualitatively we can assume near the surface for $z>0$, $m$ is positive, and for $z<0$, $m$ is negative. Hence, $p_z$ is not a good quantum number. For the low energy physics, \emph{i.e.} when we focus around $m\sim 0, k\sim 0$ we can safely take the continuum limit so  that $\sin p_z\rightarrow -i\partial/\partial z$. The wavefunction of the surface state is proportional to $e^{-\int_0^z m(z')dz'}$. There are four zero modes solutions:
\begin{align}
\mid p \uparrow \rangle =&F(z)(-1,0,i,0)^T, \ \ &\mid p \downarrow \rangle =&F(z)(0,1,0,i)^T \nonumber \\
\mid h \uparrow \rangle =&F(z)(-1,0,-i,0)^T, \ \ &\mid h \downarrow \rangle =&F(z)(0,1,0,-i)^T, \nonumber \\
F(z)=&\frac{e^{-\int_0^z m(z') d z'}}{N}
\end{align}
where N is a normalization coefficient,  $p$ and $h$ indicate  particle and hole parts respectively, and $\uparrow,\downarrow$ are associated with spin up and down. Thus, the projection of the bulk $H_{BdG}$ to these four modes is an effective surface Hamiltonian, which was written in Eq.\ \ref{eq:surfbdg}. We note that if the boundary condition changes ($M\rightarrow -M$), the surface Hamiltonian is the same in the similar basis of $(\Psi_{\uparrow},\Psi_{\downarrow},\Psi^\dagger_{\uparrow},\Psi^\dagger_{\downarrow})$ while  the basis wavefunctions of the four zero modes change. For the top and the bottom surfaces, the physics can be described by the same surface Hamiltonian. For convenience we use a simple lattice regularization for the numerical calculation of the 2D $H_{BdG}^{surf}$ in position space with two vortices.
For the lattice regularization we use a 2D lattice Dirac model tuned to the critical point:
\begin{widetext}
\begin{align}
H_{BdG}^{surf}&=
\sum_{r}
\left( \begin{array}{cc}
c_{r}^\dagger & c_{r}
\end{array} \right)
\left( \begin{array}{cc}
h(m) & \Delta i\sigma_y \\
-\Delta i\sigma_y & -h^*(m)
\end{array} \right)
\left( \begin{array}{c}
c_{r} \\
c_{r}^\dagger
\end{array} \right)  \nonumber \\
&+
\sum_{r,\epsilon}\left( \begin{array}{cc}
c_{r}^\dagger & c_{r}
\end{array} \right)
\left( \begin{array}{cc}
h(\epsilon)e^{i\int^{{\bf r}+\epsilon}_{{\bf r}} [\nabla \phi_A ({\bf r})-\frac{e}{\hbar c}{\bf A}(r)]\cdot d{\bf l}} & 0  \\
0 & -h^*(\epsilon)e^{-i\int^{{\bf r} +\epsilon}_{{\bf r}}  [\nabla \phi_B ({\bf r})-\frac{e}{\hbar c}{\bf A}(r)]\cdot d{\bf l}}
\end{array} \right)
\left( \begin{array}{c}
c_{r+\epsilon} \\
c_{r+\epsilon}^\dagger
\end{array} \right), \\
h(m)&=m\sigma_z, \ \ h(\epsilon)=\frac{\sigma_z+i\hat{\epsilon}\cdot \vec{\gamma}}{2} \nonumber
\end{align}
where $\vec{\gamma}=(\sigma_y,-\sigma_x)$ and the nearest neighbor hopping is described by $\epsilon=\pm \hat{x}$ and $\pm\hat{y}$. We set $m=-2$ to have a gapless two-component, 2D Dirac cone Hamiltonian when $\Delta$ vanishes. To avoid boundary effects, we use periodic boundary conditions in $(x,y).$  We used a size of the surface of $n_x\times n_y=120\times 60$. The positions of the two vortices are $(\pm 30,0)$. The surface Hamiltonian calculation shows the ratio of spin up and down in Fig.\ \ref{lambda}. The use of the lattice Hamiltonian is only valid if we are looking for low-energy properties of the spectrum, and for example, it does not satisfy the same symmetry properties under time-reversal that a true surface state Hamilotnian of a 3D topological insulator would.
\end{widetext}

\bibliography{TI}

\end{document}